# Electricity Harvested from Ambient Heat across Silicon Surface

Guo'an Tai*[1†] and Zihan Xu[2†]

[1] *The State Key Laboratory of Mechanics and Control of Mechanical Structures, Nanjing University of Aeronautics and Astronautics, Nanjing 210016, China. E-mail: taiguoan@nuaa.edu.cn*

[2] *Carbon Source Tech LLC, 518000, China*

[†]These authors contributed equally to this work.

**We report that electricity can be generated from limitless thermal motion of ions by two dimensional (2D) surface of silicon wafer at room temperature. A typical silicon device, on which asymmetric electrodes with Au and Ag thin films were fabricated, can generate a typical open-circuit voltage up to 0.40 V in 5 M $CuCl_2$ solution and an output current over 11 μA when a 25 kΩ resistor was loaded into the circuit. Positive correlation between the output current and the temperature, as well as the concentration, was observed. The maximum output current and power density are 17 μA and 8.6 μW/cm$^2$, respectively. The possibility of chemical reaction was excluded by four groups of control experiments. A possible dynamic drag mechanism was proposed to explain the experimental results. This finding further demonstrates that ambient heat in the environment can be harvested by 2D semiconductor surfaces or low dimensional materials and would contribute significantly to the research of renewable energy. However, this finding does not agree with the second law of thermal dynamics, and a lot of future work will be need to study the mechanism behind this phenomenon.**

Self-powered technologies, such as harvesting energy by piezoelectric effect based on ZnO nanowires and from flowing liquids or gases across graphene and carbon nanotubes, have attracted intensive interests.[1-7] However, the low output power, as well as the poor reliability are limiting their practical application. Ambient heat, which is a limitless energy source, presents universally as a form of kinetic energy of molecular, atom, particle, ion and electron in gas, liquid and solid states. Recently, we reported electricity generated from the thermal motion of ions in aqueous electrolyte solutions at room temperature by atomic layer 2D materials, such as graphene and reduced graphene oxide.[8,9]

Semiconductor surface, which can be reagarded as 2D material, is unstable and tends to be autocompensated because a large number of dangling bonds on the surface results in the appearance of delocalized electronic surface states and unstable Fermi level.[10-15] The cations on solid-liquid interface, which are governed by electrochemical double layer, are under thermal motion with a velocity of hundreds of meters per second at room temperature.[16-19] This unlimited thermal motion can offer the opportunity for momentum transformation between the electrons and the ions and thun excited the unstable electrons to be free electrons. Thus, it is possible to harvest energy in terms of limitless thermal motion of ions by semiconductor surface.

Here, we investigated the silicon (Si) devices with asymmetric electrodes configuration to capture such ionic thermal energy and convert it into electricity. Two kinds of metal electrodes with different work functions (Au and Ag) are adopted to achieve Ohmic and Schottky contacts in series for the rectifying function. A possible dynamic drag mechanism is proposed to interpret the energy transformation process. This finding shows potential of harvesting energy from ionic thermal motion at room temperature and will benefit the self-powered technology.

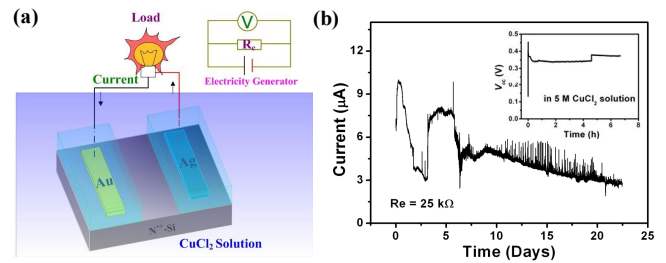

**Fig. 1**. Experimental setup and output of a Si device with Au-Ag electrodes in 5 M $CuCl_2$ solution at room temperature. (a) Schematic diagram showing the experimental setup of the device and corresponding equivalent circuit of the device. (b) Output of the device when a 25 kΩ resistor was loaded to the circuit. Inset: open-circuit voltage ($V_{oc}$) of the device.

A typical *n*-type Si wafer (marked as *n*-Si-6) with bulk and sheet carrier concentration of -9.203 × 10$^{13}$/cm$^3$ and -4.601 × 10$^{12}$/cm$^2$ was used to fabricate the device. A typical sample with the size of 1.2 × 1.2 cm$^2$, was cleaned in 5% HF for 2 mins to etch the oxidation layer away, then Au and Ag electrodes were deposited on either side of the Si surface by thermal evaporation. All the electrodes were sealed from exposing to the electrolyte solution. The exposed area was around 0.3 × 0.8 cm$^2$ between the electrodes. Fabrication processes of the device are shown in fig. S1. Then, the device was put into aqueous electrolyte solutions. The electric signals were collected by Labview program controlled KEITHLEY 2000 and 2400 multimeters. The experimental setup and the corresponding circuit were shown in Fig. 1a.. Top right view is the corresponding equivalent electrical circuit.

The open-circuit voltage ($V_{oc}$) of Si device with Au-Ag electrodes was typically up to 0.40 V in 5 M $CuCl_2$ solution (inset of Fig. 1b); the output current of this device varies from 11 to 2.7 µA and retains this value over 22 days when a 25 kΩ resistor was loaded to the circuit during all the measurements (Fig. 1b). The downtrend of the output current could be ascribed to the oxidation of the Si surface, which prevents the interaction between the cations and Si surface.[20] To verify this, we etched the exposed area of the device by 2% HF solution for 5 seconds. The result showed that the output current got back to 5 µA (Fig. S3). Another ten devices were also measured in 5 M $CuCl_2$ solution at room temperature (Fig. S4). The output current of the devices varied from 4.6 to 12.6 µA.

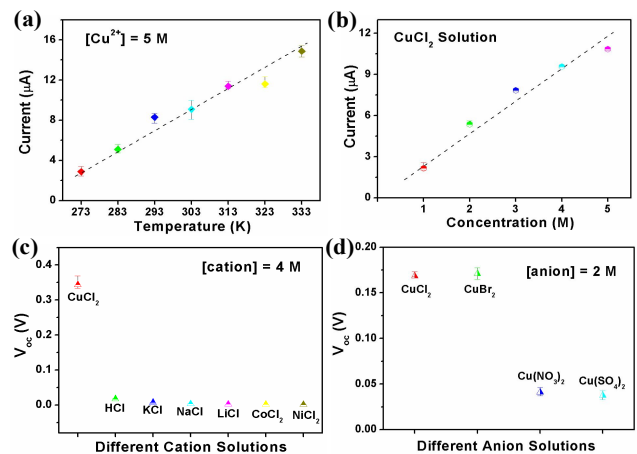

Fig. 2. Experimental measurements of the Si device. (A) Output current versus temperature relation of the Si device with Au-Ag electrodes in 5 M $CuCl_2$ solution at room temperature. (B) Output current versus concentration relation of the Si device. (C) $V_{oc}$ of devices measured in 4 M cation solutions. (D) $V_{oc}$ of devices measured in 2 M anion solutions.

The effect of temperature and concentration of ionic solutions on the output of the device were also investigated. The output current rose with the increase of temperature in 5 M $CuCl_2$ solution as indicated in Fig. 2a. The output current in 5 M $CuCl_2$ was more than 4 times higher than that in the dilute one (1 M) (Fig. 2b). A positive relationship between current and temperature, as well as concentration, was observed. Different cations (4 M) and anions (2 M) were also studied at room temperature. It is shown that $Cu^{2+}$ can induce much higher $V_{oc}$ compared with $Co^{2+}$, $Ni^{2+}$, $H^+$, $Li^+$, $Na^+$ and $K^+$ (Fig. 2c). The $V_{oc}$ measured for different anions such as $Cl^-$, $Br^-$, $NO_3^{2-}$ and $SO_4^{2-}$ has a negative correlation with the anion radii (Fig. 2d).

The possibility of chemical reaction between solutions and Si or electrodes could be reasonably excluded: (1) the cyclic voltammetry measurements (Fig. S5) show that no oxidation and reduction peaks in the C-V curve, which means no chemical reaction occurs in the system. It is shown that no electrochemical reactions occur in the potential window of the experiments;[21] (2) the $CuCl_2$ solution, in which the cell was kept for 22 days, was analysed by the inductively coupled plasma optical emission spectrometry (ICP-OES). The content of Si in the solution is 0.79 µg/mL, which is the same order of the initial one (0.1 µg/mL); (3) the Ag and Au films still intact when the device was dipped into the solution and measured after 22 days. (4) we replaced the Si wafer with glass. The result showed that the near-zero output was observed and can be regarded as noise (Fig. S6). This experiment show that the semiconductor surface play a key role in this experiment; Both (1) and (2) show that the possibility of chemical reaction between Si and the solutions can be excluded. Both (3) and (4) guarantees that no chemical reaction takes place between the electrodes and the solutions. Morever, the output was kept the same value when the sample was placed in the darkness, which mean the output energy was not arised from the light.

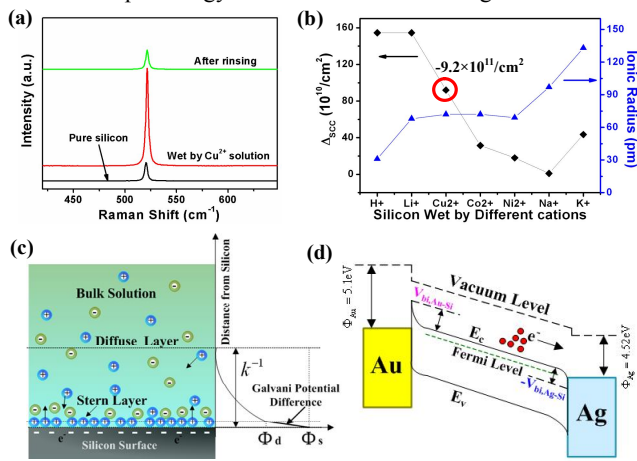

Fig. 3. Thermal ion-Si surface interaction mechanism. (A) Raman spectra of pure Si, the sample wet by 4 M $CuCl_2$ solution and rinsed by DI water. (B) Sheet carrier concentration change ($\Delta_{SCC}$) and ionic radius relation of the Si wafer when it's wet by different cations. (C) Schematic diagram of electrical double layer on Si surface in the electrolyte solutions. (D) Band diagram related to the induced electricity from the Si device. The band diagram includes band bending and carrier output mechanism in the case of spontaneous procedure due to thermal motion of ions in the solutions.

Raman spectroscopy and Hall Effect Measurements were adopted to detect carrier concentration change of Si surface after and before it was wet by the electrolyte solutions (Fig. 3a and b). After the Si was wet by $CuCl_2$ solution (middle trace, Fig. 3a), two characteristics were observed when comparing with the Raman spectrum of the pure Si: (1) Raman upshift of 1 $cm^{-1}$ was observed. (2) Raman intensity was enhanced more than 5 times. Both (1) and (2) were resulted from the carrier concentration change of the Si surface, similar to that of graphene and carbon nanotubes.[23-25] After rinsing the sample with distilled water, Raman spectrum was restored to the original state (top trace, Fig. 3a). The same phenomenon was also found in other electrolyte solutions, such as $Ni^{2+}$, $Co^{2+}$, $H^+$, $Li^+$, $Na^+$ and $K^+$ (Fig. S7). The obtained sheet carrier concentration by Hall Effect Measurements showed that all the cations can remarkably change the carrier concentration of Si surface. The results show that the larger the ionic radius, the less the carrier concentration change of the Si surface will be (Fig. 3b and Tab.S1). Importantly, negative sheet carrier concentration change up to $-9.2×10^{11}/cm^2$ after the Si wafer was wet by $Cu^{2+}$ solution shows that $Cu^{2+}$ can easily induce the excess negative charge (electron) close to Si surface, consistent with the maximum output current for $Cu^{2+}$ solution (Fig. 2c). In contrast, the induced current from the Si device in pure water is relatively small due to large molecular radius and charge neutrality, as shown in Fig. S8.

A possible dynamical drag mechanism based on electrical double layer effect was proposed to interpret the experimental results. A dynamical electric double layer is formed on the surface when the Si surface contacts with electrolyte solution, as showed in Fig. 3c. It consists of two parallel layers of charge: the first layer is the surface positive charge (cation); the second layer is negative charge (anion). This arrangement creates an electrical potential difference, called the Galvani potential difference.[26] The second layer is called the diffuse layer because it forms from free ions in the solution under the influence of stochastic thermal motion. The Debye length is the distance between Si surface and the boundary of diffuse layer, namely the thickness of electrical double layer. The Debye length can be expressed by [17, 27]

$$\kappa^{-1} = \sqrt{\frac{\varepsilon_r \varepsilon_0 k_B T}{2 N_A e^2 I}} \quad (1)$$

Where $\kappa^{-1}$, $\varepsilon_r$, $\varepsilon_0$, $k_B$, $T$, $N_A$ $e$, and $I$ are are the Debye length, the dielectric constant of the solution, the permittivity of free space, Boltzmann constant, the absolute temperature in Kelvin's, the Avogadro number, the elementary charge and the ionic strength of the electrolyte. Here,

$$I = \frac{1}{2}\sum_{i=1}^{n} c_i z_i^2 \quad (2)$$

Thus, the higher the concentration of electrolyte solutions, the smaller the Debye length will be. For example, for mono-valent NaCl (4 M) solution, $k^{-1} = 3.04/\sqrt{c}$ Å = 152 pm which is far below its ionic radius (1102 pm) and lower than the H$^+$'s ionic radius (154 pm). Moreover, the higher the valence of ions, the smaller the Debye length will be. Additionally, the number of the solvating molecules is relatively small in comparison with free ions at high concentration of electrolyte solution. Therefore, we only consider the effect of cations in the stern layer on the Si surface. Under the influence of thermal motion of ions, the thermal cations collide with the Si surface ceaselessly, analogous to a dynamic adsorption and desorption behavior which accompanied by dynamic charge (electron) drag to its surface from bulk Si due to electrostastic interaction. This phenomenon could be ascribed to auto-compensated effect of Si surface.[13] During the unlimited collide processes, the energy was transformed from the thermal cations to the surface of the Si and excited the delocalized electrons.

Furthermore, an asymmetric electrode configuration was introduced to promote the flow of random excited electrons from the Si surface and fix the current direction. The energy level is shown in Fig. 3d. Here, the work functions of Au and Ag are 5.1 and 4.52 eV, respectively; otherwise, the work function of Si wafer is 4.60 eV.[29,30] Thereby, when the devices were fabricated by the configuration, it exhibits diode characteristic in air, as shown in Fig. S2 (black line). A stronger diode behavior was observed when the device was dipped into the 5 M CuCl$_2$ solution, as indicated in Fig. S2 (red line). This leads to the appearance of build-in potential difference across the device. Since the mobility of the $n$-Si-6 wafer (~1460 cm$^2$V$^{-1}$s$^{-1}$) (Tab. S2) is much higher than that of the solution (in magnitude of 10$^{-4}$ cm$^2$V$^{-1}$s$^{-1}$), the induced electrons prefer to travel across the Si surface to the electrode instead of going into the electrolyte solution. Thus, the existence of thermal motion of ions and the build-in potential difference bring the electricity generation. In contrast, two devices with identical electrodes, namely Au-Au and Ag-Ag, were also fabricated. The current was as low as 0.1 and 2 μA for Au-Au and Ag-Ag electrodes respectively (Fig. S9). Besides, it was difficult to control the current direction, similar to the finding of our graphene devices.[8]

The proposed physical mechanism can explain the influence of different conditions on the electricity generation. The higher the temperature of Cu$^{2+}$ solution, the faster the velocity of Cu$^{2+}$, the more the excited free charge carriers and the higher the induced current will be (Fig. 2a); the higher the concentration of Cu$^{2+}$, the higher the density of Cu$^{2+}$ on Si surface, the more the electrons induced out of Si, and the higher the output current will be (Fig. 2b). The exact difference should be ascribed to the difference of ionic radii, valence electrons number n, ion mass, and so on.

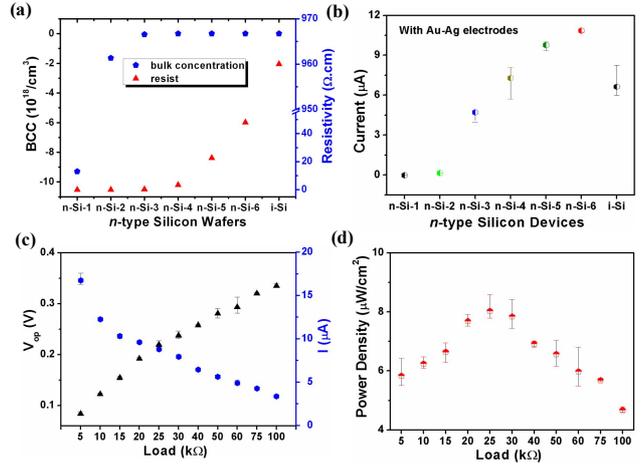

**Fig. 4**. Output of different $n$-type Si devices with Au-Ag electrodes in 5 M CuCl$_2$ at room temperature. (A) Bulk carrier concentration and resistivity. (B) Output current. (C) Output versus load resistance dependence of the $n$-Si-6 device. (D) Power density versus load resistance dependence of the device.

Different $n$-type Si wafers were also studied to investigate the relationship of their physical parameters and induced electricity. The wafers were marked as n-Si-1, n-Si-2, n-Si-3, n-Si-4, n-Si-5, n-Si-6 and i-Si, where the i-Si means intrinsic Si. The bulk carrier concentration and resisitivity by Hall Effect Measurements varies from -6.0 × 10$^{12}$ to -9.2 × 10$^{18}$/cm$^3$ and from 960 to 0.00277 Ω·cm, respectively (Fig. 4a and Tab. S2). The corresponding induced current varies from 0 to 11 μA (Fig. 4b and Tab. S2). The $p$-type Si devices also remain the same law as the $n$-type ones (Fig. S10). This demonstrates that the higher the doping concentration, the faster the recombination of the induced electron and hole on Si surface,[31,32] and the smaller the output current will be. Finally, we tested the induced voltage, output current and power density dependence of load resistance for the $n$-Si-6 Si device in the 5 M CuCl$_2$ solution. It was observed that the highest power density up to 8.6 μW/cm$^2$ was obtained from the device in our device without optimization.

## Conclusions

We have studied the Si devices to generate electricity from thermal motion of ions in aqueous electrolyte solutions at room temperature. The asymmetric electrodes pairs were deposited on either side of Si wafer to provide a potential difference acrocss the wafer surface, which can drive the random electrons to travel with the fixed direction. A dynamic drag mechanism was proposed to interpret the experimental results. Temperature and concentration dependence of the induced current support the proposed mechanism. This finding present here opens the door for the development of self-powered technology to harvest energy from the ambient heat. On the basis of the experiments, we predicat that all the low-dimensional materials, as well as the surface of all the semiconductors should be used to generate electricity from the amibient heat with the help of asymmetrical electrodes.

However,, this finding does not agree with the second law of thermodynamics, which limits the utilization of the random thermal motion of ions to be spontaneously collected to produce electricity. We cannot explain why either this experiment or the previous experiment of graphene did not agree with the traditional theory. More research will be required to fully understand this phenomenon.

## Acknowledgments

We thank Prof. H. X. Chang and Mr Z. K. Liu for fruitful discussions. This work was supported by National and Jiangsu Province Postdoctoral Science Foundation (201003582, 20090451207, 0901073C), Jiangsu Province NSF (BK2010501), Ph.D. PFMEC (20103218120035) and Innovation Fund of NUAA (NS2012009).

## References

† Electronic Supplementary Information (ESI) available: [Experimental methods, fabrication process of the device, control experiments].

Supporting information for

# Electricity Harvested from Ambient Heat across Silicon Surface

Guo'an Tai*[1†] and Zihan Xu[2†]

[1] The State Key Laboratory of Mechanics and Control of Mechanical Structures, Nanjing University of Aeronautics and Astronautics, Nanjing 210016, China.

[2] Carbon Source Tech. LLC 518000, China

Correspondence to: taiguoan@nuaa.edu.cn

## 1. Materials and Methods

Silicon (Si) wafers with different type and doping concentration were used to investigate the electricity generation. The wafers was cut into small pieces of 1.2 × 1.2 cm$^2$. Then they were etched by 2% HF solutions and rinsed by DI water many times. The devices fabricated have the Au/Si/Ag structure, where the Au films with thickness of around 150 nm were produced by thermal evaporation and annealed at 400 °C for 5 min in N$_2$ environment to enhance the Ohmic contact. Then, Ag thin films were also deposited by thermal evaporation and annealed at 120 °C for 15 min to enhance the Schottky contact. The distance between both electrodes is 0.5 cm and their lengths and widths are 0.8 and 0.3 cm, respectively, which were patterned by a metal mask. The devices were encapsulated by clear paste to guarantee that the electrical performance is not influenced by the solution constituents.

The crystal structure of silicon was examined by Raman spectroscopy (Renishaw (inVia)) with excited wavelength of 633 nm. The mobilities and carrier concentration change of the Si surfaces were measured by Hall Effect Measurement System (Ecopia, HMS-5000). For performing the measurements, the Au dots with 0.2 mm diameter and 150 nm thickness were deposited on four sides of the Si wafers of 1.2 × 1.2 cm$^2$ and then they were annealed at 400 $^o$C for 5 min to create the Ohmic contact. The content of Si dissolved in the $CuCl_2$ solution was detected by inductively coupled plasma optical emission spectrometry (ICP-OES). The Cyclic Voltammetry measurements were a Potentiostatic Electrochemical Station (CHI 660C) with a sweep rate of 100 mV/s in the potential range of -1.0 to 1.0 V. Saturated calomel electrode (SCE), silicon wafer and platinum foil were adopted as a reference electrode, working electrode and counter electrode, respectively. KEITHLEY 2000 programmable electrometer was used to measure the open-circuit and output voltage of the devices. KEITHLEY 2400 programmable electrometer was applied to collect current-voltage characteristics of the devices.

## 2. Figures S1~S10 and Tables S1-S2

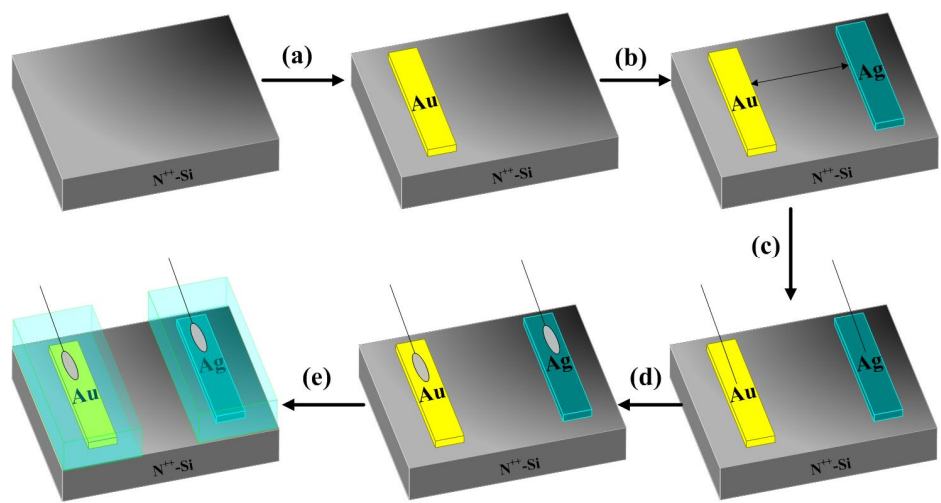

**Figure S1.** Schematic diagram of fabrication processes for Si devices. (a) Etching of Si wafer and deposition of Au thin film on the surface by thermal evaporation. (b) Annealing of Au thin film and deposition of Ag thin film on the surface by thermal evaporation. (c) and (d) Connecting Ag coated copper wires with the thin films by conductive gel. (e) Sealing the electrodes with clear paste.

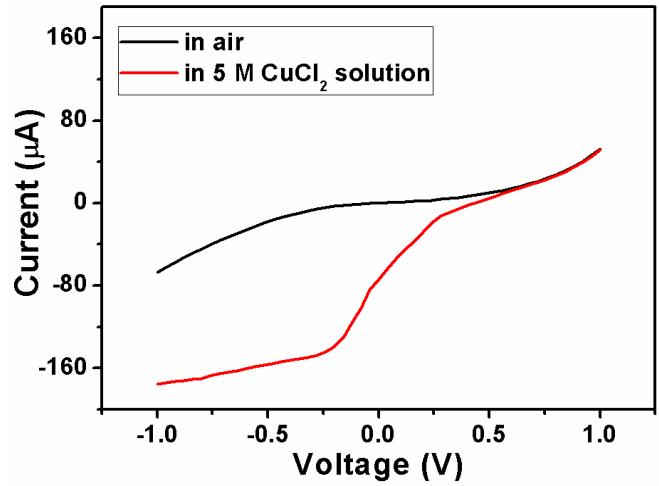

**Figure S2.** Current-voltage (*I-V*) characteristics of the silicon device with Au-Ag electrodes before (dark line) and after (red line) it was dipped into 5 M $CuCl_2$ solution.

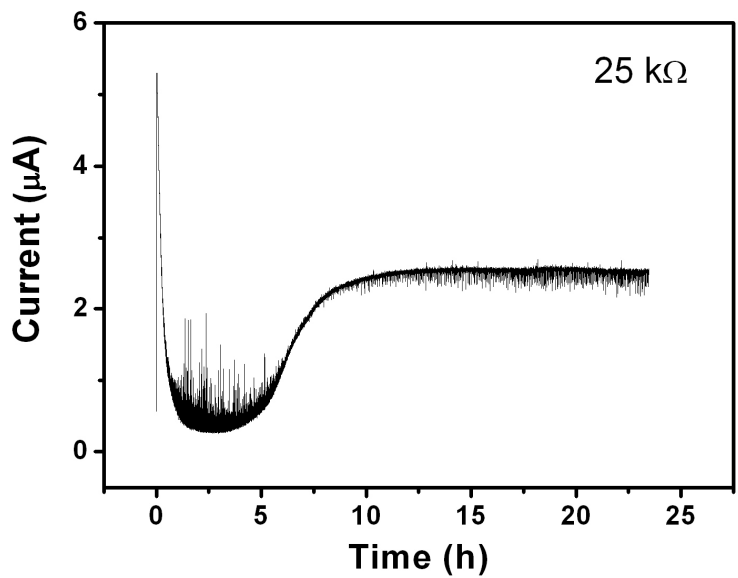

**Figure S3.** Output of the Si device with Au-Ag electrodes after it was dipped into 5 M $CuCl_2$ solution again.

This downtrend may be ascribed to poor contact between electrodes and Si surface arising from swelling of clear paste in the solution for a long time.

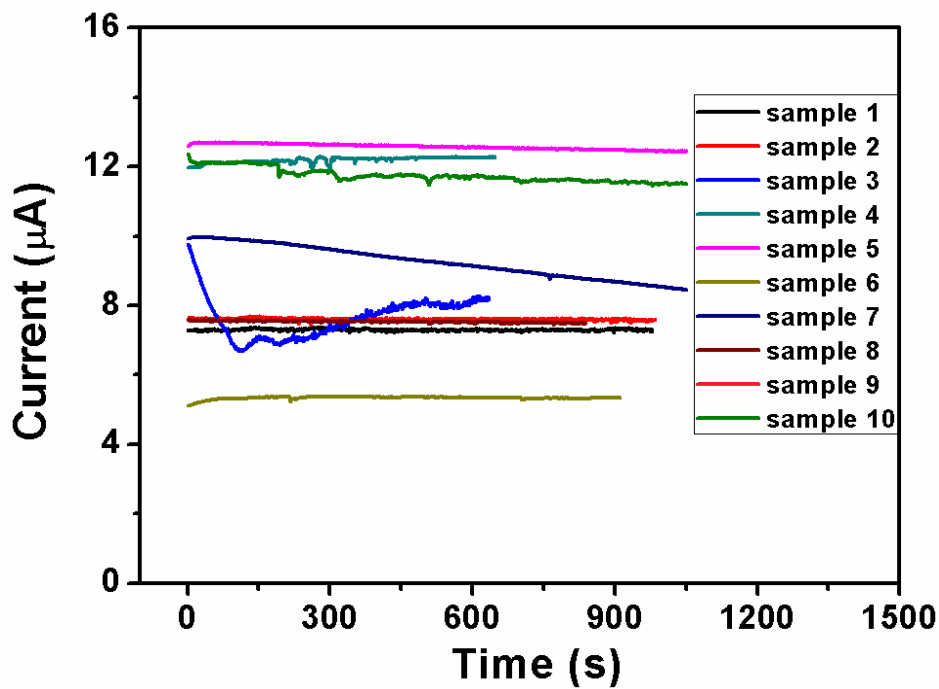

**Figure S4**. Output of ten Si devices in 5 M CuCl$_2$ solution at room temperature when a 25 kΩ resistor was loaded to the circuit.

The measured current varies from 5.35 to 11.74 μA. The variation depends upon the quality of deposition and treatment of asymmetric electrodes on Si wafers. We found that the better the Schottky diode in air, the higher the output will be.

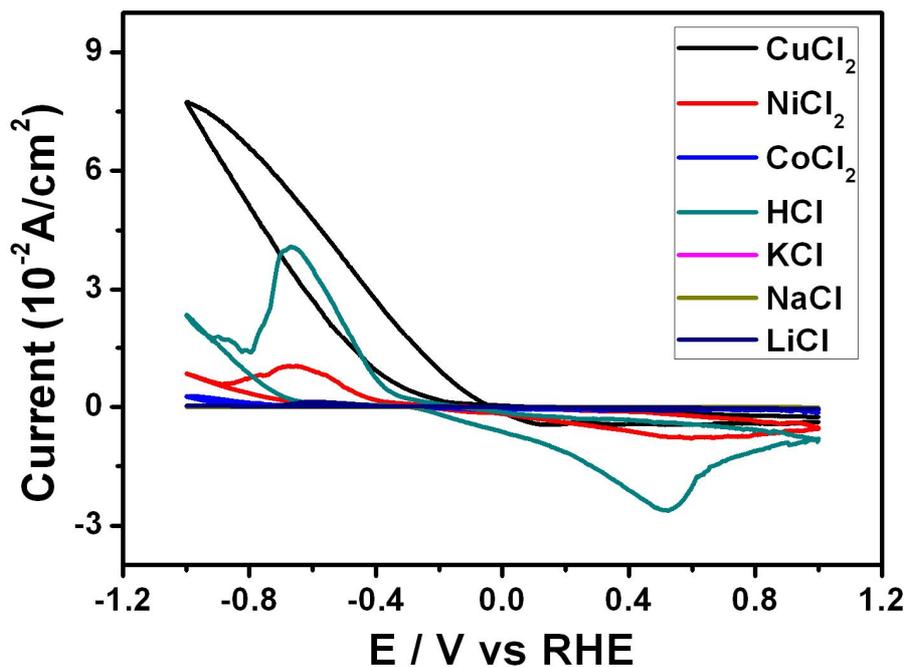

**Figure S5**. Cyclic voltammogram of the *n*-Si-6 wafer in different cation solutions. The concentration of the solutions was 4 M and scan rate is 100 mV/s.

As can be seen from Figure S5, no oxidation and reduction peaks occur between the wafer and $CuCl_2$ solution in its potential window, which means that no chemical reaction takes place between the wafer and the $CuCl_2$ solution. Nevertheless, strong oxidation and reduction peaks appear between the wafer and HCl solution, between it and $NiCl_2$ solution, between it and $CoCl_2$ solution or between it and LiCl solution. No these characteristic peaks occur between the wafer and the NaCl solution, as well as between it and the KCl solution. As can be observed from Table S1, the open-circuit voltage is relatively low when the device dipped in HCl, $NiCl_2$, $CoCl_2$, LiCl, NaCl and KCl solutions as compared with that in $CuCl_2$ solution. These results demonstrated that the possibility of chemical reaction between the wafer and the solutions could be reasonably excluded.

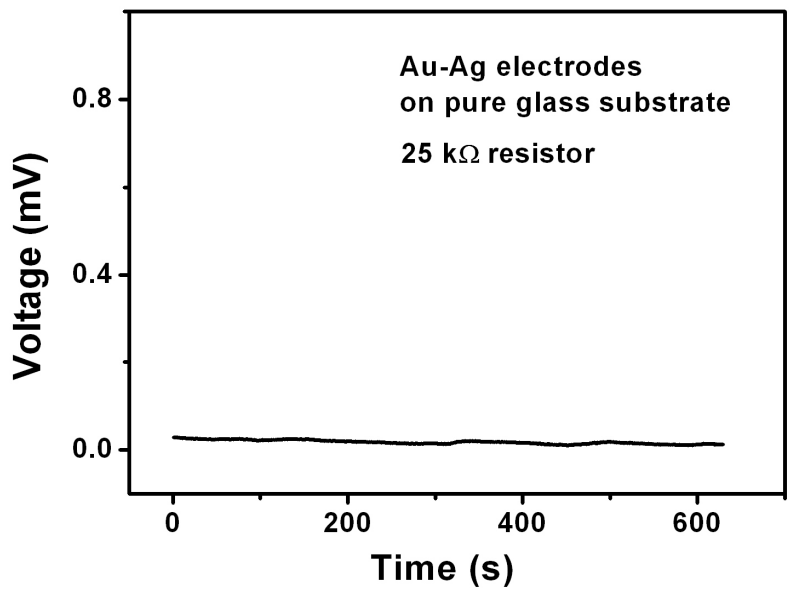

**Figure S6**. $V_{op}$ versus time relation of a device with Au-Ag electrodes on pure glass substrate.

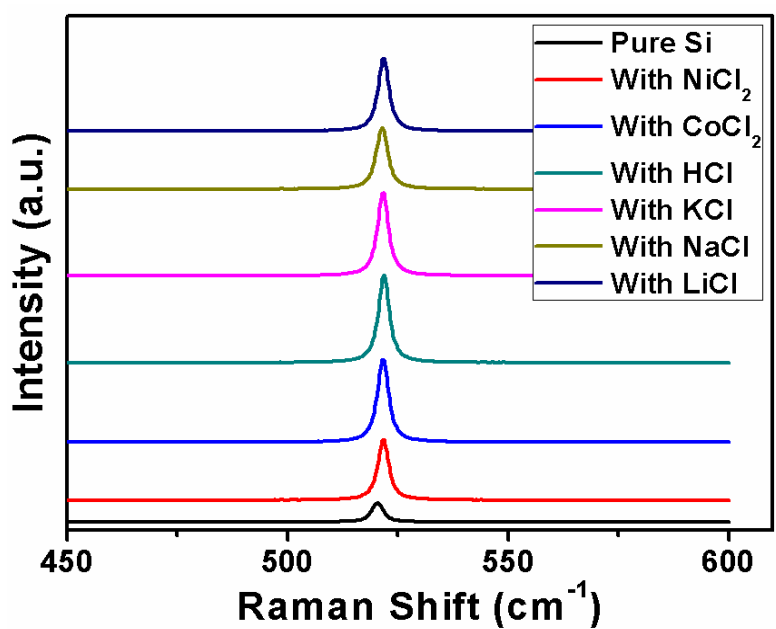

**Figure S7**. Raman spectra of pure Si and samples wet by different solutions: $NiCl_2$, $CoCl_2$, HCl, KCl, NaCl and LiCl.

As can be seen from Raman spectrum of pure Si (bottom, Fig. S7), the one band located at ~520.2 cm$^{-1}$, which is considered as the characteristic band of Si. Raman upshifts up to 1 cm$^{-1}$ and enhancement of Raman intensities were observed for the n-Si-6 wafer after it is dipped in all the solutions.

**Table S1.** Ionic type, ionic radius, sheet carrier concentration (SCC, $\times 10^{12}/cm^2$), change of sheet carrier concentration ($\Delta_{SCC}$, $\times 10^{12}/cm^2$) of the Si wafer wet by different cation solutions as compared to pure Si. The corresponding open-circuit voltage of the device dipped into different electrolyte solutions.

| Ionic Type | $H^+$ | $Li^+$ | $Cu^{2+}$ | $Co^{2+}$ | $Ni^{2+}$ | $Na^+$ | $K^+$ |
|---|---|---|---|---|---|---|---|
| Ionic Radius (pm) | 31 | 68 | 72 | 72 | 69 | 97 | 133 |
| SCC ($\times 10^{12}/cm^2$) | -4.447 | -4.447 | -4.693 | -4.57 | -4.583 | -4.6 | -4.548 |
| $\Delta_{CC}(\times 10^{10}/cm^2)$ | 154.5 | 154.5 | -92 | 31.5 | 18 | 1.0 | 43.5 |
| $V_{oc}(V)$ | 0.018 | 0.002 | 0.341 | 0.002 | 0.001 | 0.003 | 0.008 |

Note: (1) The SCC of pure Si is $-4.601 \times 10^{12}/cm^2$.

(2) The SCCs provided here is an average value of the measurements for ten times.

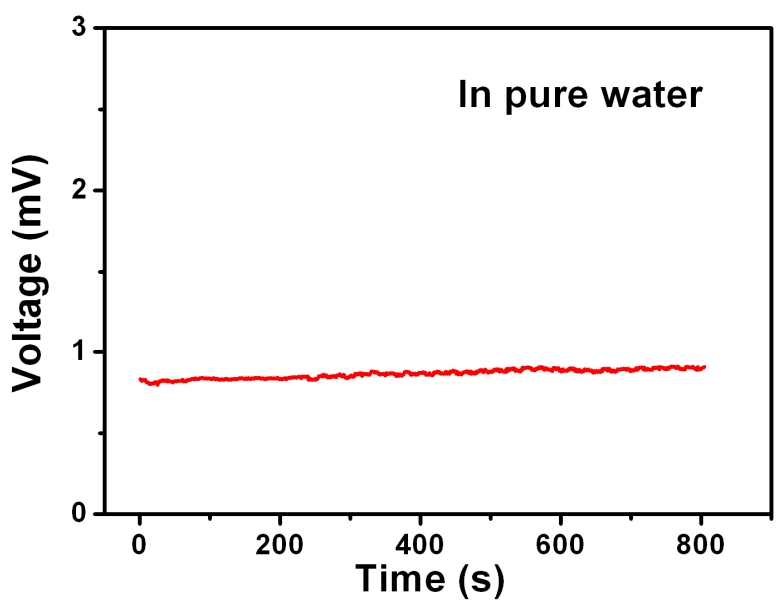

**Figure S8.** $V_{op}$ of a n-Si-6 device in pure water with a 25 kΩ resistor loaded to the circuit.

**Table S2**. The Hall mobilities, resistivities, bulk carrier concentration (BCC), sheet carrier concentration (SCC) and induced output of different Si wafers. The induced output was obtained by loaded a 25 kΩ resistor to the circuit.

| Different type's Si | Crystal Orientation | Measured Parameters | | | | |
|---|---|---|---|---|---|---|
| | | Hall Mobility ($cm^2/V·s$) | Resistivity ($\Omega·cm$) | BCC ($/cm^3$) | SCC ($/cm^2$) | Output Current ($\mu A$) |
| n-Si-1 | (111) | 327.1 | 2.77E-3 | -9.30E18 | -4.65E17 | 4.08E-4 |
| n-Si-2 | (100) | 686.4 | 2.04E-3 | -1.65E18 | -8.23E16 | 0.14 |
| n-Si-3 | (100) | 666.5 | 0.191 | -4.89E16 | -2.45E15 | 4.74 |
| n-Si-4 | (100) | 1358.5 | 3.385 | -1.36E15 | -3.31E13 | 7.36 |
| n-Si-5 | (100) | 712.9 | 22.545 | -3.90E14 | -1.95E13 | 9.84 |
| n-Si-6 | (111) | 1429.5 | 47.86 | -9.13E13 | -4.56E12 | 10.86 |
| p-Si-1 | (100) | 256.7 | 3.91E-3 | 3.85E18 | 1.92E17 | 7.45E-4 |
| p-Si-2 | (100) | 126.2 | 1.35E-2 | 3.14E18 | 1.57E17 | 4.99E-3 |
| p-Si-3 | (100) | 306.9 | 3.872 | 5.26E15 | 1.05E14 | 4.46 |
| p-Si-4 | (100) | 304.9 | 4.164 | 4.91E15 | 9.83E14 | 7.35 |
| i-Si | (100) | 1082 | 960.0 | -6.02E12 | -3.01E11 | 6.45 |

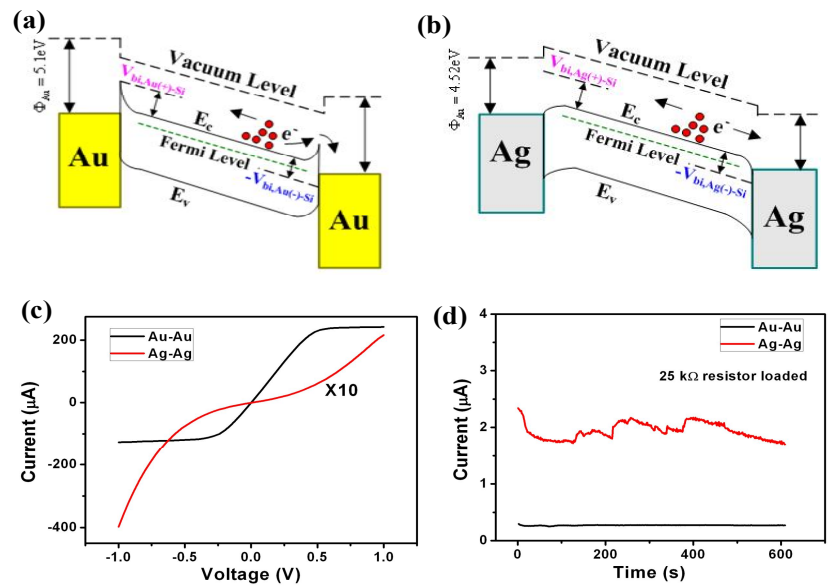

**Figure S9**. Work function tuning mechanism. (a) Energy level diagram for the Si contacted with Au-Au electrodes. (b) Energy level diagram for the Si contacted with Ag-Ag electrodes. The green dashed line denotes the Fermi level, the black solid line represents the energy of silicon at valence and conduction bands, the yellow rectangle is the Au contact, and the silvery white rectangle is the

Ag contact. $\phi_{Au}$ and $\phi_{Ag}$ are the work function of Au and Ag, respectively. $V_{bi,Au-Si}$ and $V_{bi,Ag-Si}$ are the built-in potential between Au and Si and between Au and Si. (c) Current-voltage characteristic of the n-Si-6 device with Au-Au or Ag-Ag electrodes. (d) Output of the n-Si-6 device with Au-Au or Ag-Ag electrodes.

The current direction can be determined for the device with Au-Ag as asymmetric electrodes. This is because work function of gold (5.1 eV) is larger than that of Si (4.60 eV), while work function (4.52 eV) of silver is smaller than that of the Si. From Figure S9d, we can observe that the relatively small output can be obtained by the device with symmetric electrodes. The output up to 2 µA for the device with Ag-Ag electrodes is owing to the near work function between Ag and Si, while only 0.2 µA output current was achieved when Au-Au electrodes were adopted.

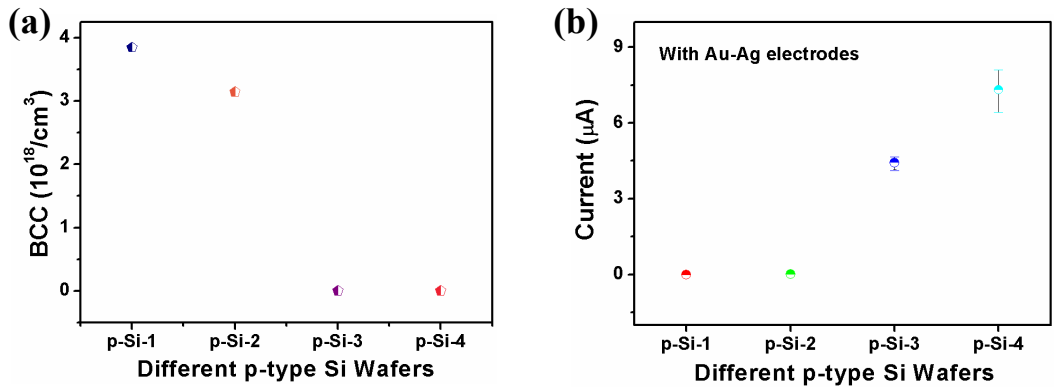

**Figure S10.** Output of different *p*-type Si devices with Au-Ag electrodes in 5 M $CuCl_2$ at room temperature. (a) Bulk carrier concentration. (b) Output current.